\documentstyle[graphicx, aps]{revtex}
\begin{document}

\title{Long-range interaction of two metastable rare-gas atoms}
\author{A. Derevianko and A. Dalgarno}
\address{Institute for Theoretical Atomic and Molecular Physics,\\
Harvard-Smithsonian Center for Astrophysics,
Cambridge, Massachusetts 02138}
\date{\today}
\maketitle
\begin{abstract}
We present semiempirical calculations of long-range van der Waals interactions
for two interacting metastable rare-gas atoms Ne through Xe. 
Dispersion coefficients $C_6$ 
are obtained for homonuclear molecular potentials asymptotically connecting to the 
$ns(3/2)_{2} + ns(3/2)_{2}$ atomic states.
The estimated uncertainty of the calculated $C_6$ dispersion
coefficients is 4\%. 
\end{abstract}
\pacs{PACS: 34.20-b, 32.10.Dk, 32.70.Cs}

\maketitle

Motivated by cold-collision studies of metastable 
rare-gas atoms~\cite{Orzel_Xe_PRA99,Doery_sp_NeKr_98,Doery98_ssC6,Optical_Control_Xe_PRL95} 
and prospects of achieving Bose-Einstein Condensation in these
systems~\cite{Beijerinck_PRA00}, we present calculations of long-range
dispersion (van der Waals) coefficients for two atoms interacting in the  
$ns(3/2)_{2}$ atomic 
states ($n=3$ for Ne, $n=4$ for Ar, $n=5$ for Kr, and $n=6$ for Xe).
The metastable states
have long lifetimes, 43 sec for Xe~\cite{Xe_lifetime},  
decaying to the ground $\,^1\!S_0$ state 
by a weak magnetic-quadrupole  transition. With such a long lifetime 
the metastable atom behaves as an effective ground state in 
experiments.
Compared to  alkali-metal systems,
an attractive feature of the noble gas atoms is the availability
of isotopes with zero nuclear spin. The lack of hyperfine structure
leads to a substantial simplification of molecular potentials,
though some complexity arises due to the nonvanishing
total electron angular momentum ($J$=2) of the metastable state.
The anisotropy leads to fifteen distinct long-range
molecular states connecting to the $ns(3/2)_{2} + ns(3/2)_{2}$ asymptotic configuration.

Our theoretical treatment of long-range interactions is similar to  recent 
high-precision calculations of 
van der Waals coefficients for alkali-metal atoms~\cite{DereviankoC6_PRL99}.
By using many-body methods and accurate experimental
matrix elements for the principal transitions,  leading dispersion coefficients $C_6$ 
were determined to an accuracy  better than 1\% for Na, K, and Rb,
and of 1\% for Cs and 1.5\% for Fr. 
The {\em semiempirical} values of $C_6$ coefficients for metastable noble-gas
atoms obtained here have an estimated uncertainty of 4\%.
The approach relies on the determination of dynamic polarizability functions.
To construct the polarizabilities
we combine experimental 
lifetime~
\cite{Kandela84,Hartmetz,Volz98,SchmoranzerVolz_Kr93,%
Inoue_Xe_JCP84,Husson_OptCom72,Allen_JOSA69,Wiese89} 
and energy  data of the excited states with accurate semiempirical 
dynamic polarizabilities of the ground states of noble-gas atoms~\cite{KumarMeath85}. 
The theoretical lifetimes and branching ratios~\cite{Aymar_78} are adjusted
to reproduce the measured static polarizabilities~\cite{Molof74}, 
which are known with a 2\% uncertainty.
We estimated the additional small contributions within the Dirac-Hartree-Fock framework.  
The resulting polarizabilities satisfy the Thomas-Reiche-Kuhn oscillator strength sum rule. 

The Racah notation for atomic levels is used. 
The particle-hole states are labeled as $n\ell (K)_J$ or  $n\ell'(K)_J$, 
where $n$ and $\ell$ are the principal and the orbital 
angular momentum quantum numbers  of the valence electron and  
${\mathbf K}={\mathbf J}_c+{\mathbf \ell}$, where
$J_c$ is the angular momentum of the core. 
The primed configurations converge to a Rydberg series 
limit with a hole in the $(n-1)p_{1/2}$
state, and the unprimed to a hole in the $(n-1)p_{3/2}$ state.
The manifold
of the lowest $ns$ valence states has four fine-structure
states $ns'(1/2)_{0,1}$ and $ns(3/2)_{1,2}$, and the lowest 
$np$ manifold consists of ten states. 
We investigate here the molecular potentials 
asymptotically connecting to the  $ns(3/2)_{2}$ atomic states. 

We calculate the long-range molecular potentials in the framework
of  Rayleigh-Schroedinger perturbation theory.
The  basis
functions are defined as products of atomic wavefunctions
\begin{equation}
|M_{1}M_{2};\Omega \rangle =|ns(3/2)_{2}M_{1}\rangle
_{1}|ns(3/2)_{2}M_{2}\rangle _{2}\,,  \label{Eqn_basUncoupled}
\label{Eqbasis}
\end{equation}
where the index 1(2) describes the wavefunction
located on the center 1(2) and $\Omega =M_{1}+M_{2}$, $M_{1,2}$ being projections of the
atomic total angular momentum on the internuclear axis. Due to the axial symmetry
of a dimer 
$\Omega$ is a conserved quantum number. It takes values ranging from zero to four.
The two-atom basis~(\ref{Eqbasis}) is degenerate and the correct molecular wavefunctions
are obtained by diagonalizing the molecular Hamiltonian
\begin{equation}
\hat{H}=\hat{H}_{1}+\hat{H}_{2}+\hat{V}(R) \label{Eq_Hamilt}
\,.
\end{equation}
In expression~(\ref{Eq_Hamilt}) $\hat{H}_{k}$ represent the Hamiltonians 
of the two non-interacting atoms,
and $\hat{V}(R)$ is the interaction potential at an internuclear distance $R$. 
The energy of the $ns(3/2)_{2}$ metastable state 
is designated as ${\mathcal E}^*$.
Then in the model space~(\ref{Eqbasis})
\[
 \left(\hat{H}_{1}+\hat{H}_{2} \right) |M_{1}M_{2};\Omega \rangle =
 2 {\mathcal E}^* \, |M_{1}M_{2};\Omega \rangle \, .
\]
The residual electrostatic potential $\hat{V}(R)$ is
defined as the full Coulomb interaction energy
in the dimer excluding interactions of the atomic electrons with 
their parent nuclei.  
 
The multipole interactions ($L=1$ for dipole-dipole, and
 $L=2$ for quadrupole-quadrupole interactions) are given by~\cite{DalgarnoDavison66}
\begin{equation}
V_{LL}(R)= \frac{1}{R^{2L+1}} 
\sum_{\mu =-L}^{L}\frac{(2L)!}{(L-\mu )!(L+\mu )!} 
\left(T_{\mu}^{(L)}\right)_1  \left(T_{-\mu }^{(L)}\right)_2\, ,
\end{equation}
with the multipole spherical tensors
\begin{equation}
T_{\mu }^{(L)}=-|e|\sum_{i}r_{i}^{L}C_{\mu }^{(L)}(\hat{\mathbf{r}}_{i}) \, ,
\end{equation}
where the summation is over atomic electrons, 
${\mathbf r}_i$ is the position vector of electron $i$, 
and $C_{\mu }^{(L)}(\hat{\mathbf{r}}_{i})$
are reduced spherical harmonics~\cite{Varshalovich}.
In the following we write $d_\mu = T_{\mu }^{(1)}$ and $Q_\mu = T_{\mu }^{(2)}$.

The lowest-order 
contribution to the term energies arises from the quadrupole-quadrupole 
interaction $\hat{V}_{qq}$, which varies as $1/R^5$.
However, the corresponding $C_5$ coefficients are only of the 
order $10^1-10^2$ a.u.~\cite{Doery_sp_NeKr_98}, and
the dominant contribution appears in the second order in  $\hat{V}(R)$,
arising from the dipole-dipole  interaction $\hat{V}_{dd}$. The second-order dipole
interaction is proportional to $1/R^6$, and the associated dispersion coefficient 
$C_6$ is of the order of $10^4-10^5$ a.u..
Applying the formalism of  
degenerate perturbation theory in second order~\cite{Dalgarno_PT}, 
we obtain an effective 
Hamiltonian within the two-atom basis Eq.~(\ref{Eqbasis})
\begin{equation}
\langle m|H_{\mathrm{eff}}^{\left( 2\right) }|n\rangle 
= 2{\mathcal E}^* \delta_{mn} 
+\langle m|\hat{V}_{qq}|n\rangle +
\sum_{\Psi _{i}}\frac{\langle m|%
\hat{V}_{dd}|\Psi_{i}\rangle \langle \Psi _{i}|\hat{V}_{dd}|n\rangle }{%
2{\mathcal E}^* - E_{i}}\,.  \label{Heff2_Eqn}
\end{equation}
The intermediate molecular state $|\Psi_{i}\rangle$ with  unperturbed energy
$E_{i}$ runs over a {\em complete} set of two-atom states, excluding the
model-space states Eq.~(\ref{Eqbasis}). The formalism
of the generalized Bloch equation~\cite{LindgrenMorrison} would allow the inclusion in the model space
of the other three atomic states in the $ns$
manifold and would account for the mixing of the different fine-structure levels; 
but such a large model space is not necessary for $R>10$ a.u.. 
The position of the avoided level crossing can be estimated from
$R_{l.c.} \approx 
2 C_6/({\mathcal E}_{ns(3/2)_1} - {\mathcal E}_{ns(3/2)_2}) \sim 10$ a.u..
The $\Omega=4$ molecular term is unique and 
being unaffected by avoided crossings, 
the region of applicability is extended to $R \sim n$ a.u.,
before  the electronic clouds start to overlap.  
The effect of  the quadrupole-quadrupole interaction on the term energy can 
be disregarded at values of  $R \ll C_6/C_5 \sim 10^3$ a.u.. 
The quadrupole-quadrupole correction is discussed by  Doery {\em et al.}~\cite{Doery98_ssC6}.

Using the Wigner-Eckart theorem, we can represent 
the matrix element of the dipole-dipole term in the
effective second-order Hamiltonian  as
\begin{eqnarray}
\lefteqn{ \sum_{\Psi _{i}}\frac{\langle M_{1}M_{2};\Omega |\hat{V}_{dd}|\Psi
_{i}\rangle \langle \Psi _{i}|\hat{V}_{dd}|M_{1}^{\prime }M_{2}^{\prime
};\Omega \rangle }{2{\cal E}^{\ast }-E_{i}} =} \nonumber \\
&& -\frac{1}{R^6}  \sum_{I I'} C_6^{J_a J_b} (-1)^{J_a + J_b}
\frac{2}{3} \sum_{\lambda \mu} 
w_\lambda^1 w_\mu^1 {\mathcal A}^{J_a}_{\lambda \mu}( M_1, M'_1)
{\mathcal A}^{J_b}_{-\lambda -\mu}( \Omega-M_1, \Omega-M'_1) \, .
\end{eqnarray}
The dipole weights $w_\mu^1$ are   $w^1_{+1}=w^1_{-1}=1$, and $w^1_0=2$.
$J_a$ and $J_b$ are the corresponding total angular momenta of intermediate atomic states of atoms 1 and
2, and 
\begin{equation}
{\mathcal A}^{I}_{\lambda \mu}( M_1, M'_1) = 
\left( 
\begin{array}{ccc}
2 & 1 & I \\ 
-M_{1} & \mu & m
\end{array}
\right) \left( 
\begin{array}{ccc}
 I & 1 & 2 \\ 
-m & \lambda  & M_{1}^{\prime }
\end{array}
\right) \, ,
\end{equation}
where $m = \frac{1}{2} ( M_1 + M'_1 +\lambda -\mu )$. The intermediate (uncoupled)
dispersion coefficients are
\begin{equation}
C_6^{J_a J_b} = \frac{3}{\pi} \int_0^\infty 
d\omega \, S_{J_a}(i\omega) S_{J_b}(i\omega) \, .
\label{Eqn_C6int}
\end{equation}
The reduced dynamic dipole polarizability $S_{I}(i \omega)$ 
of purely imaginary argument is 
defined as the sum over atomic states $|\alpha_I I M_I \rangle$ with total 
angular momentum $I$ and
energy ${\mathcal E}_{\alpha_I}$
\begin{equation}
S_{I}(i \omega) = \sum_{\alpha_I}
\frac{({\mathcal E}^* - {\mathcal E}_{\alpha_I}) 
 \langle ns(3/2)_{2}||d||\alpha_I I \rangle \langle \alpha_I I  ||d|| ns(3/2)_{2}\rangle }
 {({\mathcal E}^* - {\mathcal E}_{\alpha_I})^2 + \omega^2 } \,.
 \label{Eqn_SIw}
\end{equation}
Here $\alpha_I$ stands for all quantum numbers of the intermediate state,
except for the total angular momentum $I$, and $\langle i||d||j\rangle$
are the reduced electric-dipole matrix elements, defined by the Wigner-Eckart theorem. 
Three sums with $I=1,2,3$ are allowed by electric-dipole
selection rules. 

We proceed now to construct the dipole polarizability functions 
$S_{I}(i \omega)$, Eq.~(\ref{Eqn_SIw}), 
evaluate the uncoupled dispersion coefficients~(\ref{Eqn_C6int}),
and set and diagonalize the second-order effective Hamiltonian~(\ref{Heff2_Eqn}). 

The functions $S_{I}(i \omega)$ satisfy several sum rules. 
In particular, the static tensor dipole polarizability $\alpha_{zz}(M)$ 
of the  $ns(3/2)_{2}$ state may be expressed as 
\begin{equation}
\alpha_{zz}(M)= -2 \, \sum_I (-1)^I 
\left( 
\begin{array}{ccc}
2 & 1 & I \\ 
-M & 0 & M
\end{array}
\right) ^{2} 
\;\times\,S_{I}(0)\,. 
\end{equation}
The static tensor dipole polarizabilities $\alpha_{zz}(M=1)$ and 
$\alpha_{zz}(M=2)$ of 
metastable noble-gas atoms Ne through Xe  have been
measured by Molof {\em et al.}~\cite{Molof74} to within an error of 2\%. 
In the present calculations the values of $S_I(0)$ are adjusted
to reproduce these experimental values.
In addition, as $\omega \rightarrow \infty$, the reduced polarizabilities
satisfy the nonrelativistic Thomas-Reiche-Kuhn (TRK) sum rule
\begin{equation}
\sum_n f_{kn}  = \frac{2}{15} \sum_I (-1)^{I+1} S_I(i \infty)  = N \, ,
\label{Eqn_TRK}
\end{equation}
$N$ being the number of electrons in the atom. Our constructed polarizabilities
satisfy the sum rule.

It is instructive to consider the action of a one-particle operator on the 
reference particle-hole
Slater determinant $ns(3/2)_{2}$ in the independent electron approximation. Such an operator can 
 (i) annihilate the reference 
particle-hole pair;
(ii) promote a valence electron
from $ns_{1/2}$ state to another valence state $mp_{1/2,3/2}$,
the state of the $(n-1)p_{3/2}$ hole being unchanged; 
(iii) deexcite the $(n-1)p_{3/2}$ hole into some other hole state,
     the valence state remaining the same, 
and 
(iv) create another particle-hole pair in addition to the reference
      pair.
According to such a classification it is convenient to break 
the polarizability function, Eq.~(\ref{Eqn_SIw}), into three 
contributions $S_I^k$
corresponding to a number $k$ of particle-hole pairs  in the intermediate
state $|\alpha_I I\rangle$
\[
  S_I = S_I^0 + S_I^1 + S_I^2 \, . 
\]
Since an electric-dipole 
transition from $ns(3/2)_{2}$ to the closed-core state $\,^1\!S_0$ is prohibited 
by the angular selection
rules, $S_I^0 \equiv 0$. 

The sum $S_I^1$ is
separated into contributions from the intermediate states in
the lowest $np$ fine-structure multiplet $\left(S_I^1\right)_{np}$,
and the rest of the sum  $\left(S_I^1\right)'$
\[
 S_I^1 = \left(S_I^1\right)_{np} + \left(S_I^1\right)'.
\]
The first term is the dominant contribution.
We calculate $\left(S_I^1\right)_{np}$  
using experimental values of transition energies and 
decay rates, and adjusted branching ratios.
The rest of the sum over valence states (including bound and continuum
states) $(S_I^1)'$  is estimated in the  
Dirac-Hartree-Fock (DHF) approximation.
The metastable state $|vh;J\rangle$ in  lowest order
is represented as a combination of the $|v\rangle=ns_{1/2}$
particle state and the hole state $|h\rangle=(n-1)p_{3/2}$, coupled to
the total angular momentum $J=2$, $[((n-1)p_{3/2})^{-1} ns_{1/2}]_2$. 
The lowest-order energy of such a state
is ${\mathcal E}^* = \varepsilon_v - \varepsilon_h$, $\varepsilon_i$ being 
the energy of the DHF orbital $|i\rangle$. The intermediate state
is represented as $|ma;I\rangle$, a particle state $m$ coupled with
hole state $a$ to the total momentum $I$.
Explicitly,
\begin{eqnarray*}
(S^1_I)'(i\omega) =  
 (-1)^{J+I} [I][J] 
\sum_{ma}{}' &&
\left(  \delta_{ha} \frac{ (\varepsilon_v - \varepsilon_m) \, \langle v || d ||m \rangle^2}
{ (\varepsilon_v - \varepsilon_m)^2 + \omega^2 } 
\left\{ 
\begin{array}{ccc}
J & 1 & I \\ 
j_m & j_h & j_v 
\end{array}
\right\}^2 
\right.  \\ &&
\left. 
+\delta_{mv} 
\frac{ (\varepsilon_a - \varepsilon_h) \, \langle h ||d|| a \rangle^2 }
     { (\varepsilon_a - \varepsilon_h)^2 + \omega^2} 
\left\{ 
\begin{array}{ccc}
J & 1 & I \\ 
j_a & j_v & j_h 
\end{array}
\right\}^2  \right) \,,
\end{eqnarray*}
where $[K] \equiv 2K+1$, the summation is performed over the core orbitals $a$ and
excited states $m$, excluding states of the lowest $np$ multiplet, and $J=2$.
The first sum is associated with excitation
of the valence electron (case (ii))  
while the second sum with deexcitation of the hole state (case (iii)). 
To arrive at this result  we disregarded the coupling between
levels within the same fine-structure multiplet. For example, for 
Ne the $[(2p_{3/2})^{-1} 4p_{1/2}]_2$, $[(2p_{3/2})^{-1} 4p_{3/2}]_2$,
and $[(2p_{1/2})^{-1} 4p_{3/2}]_2$ states are summed over
independently, even though the correct lowest-order wave-function
is a linear combination of them. This approximation 
corresponds to a disregard of the small difference between the energies
of the coupled and uncoupled states. Since the contribution $(S_I^1)'$
is relatively small, such an estimate suffices at the present
level of accuracy.  Numerical evaluation of $(S_I^1)'$ has been performed using
a B-spline basis set in the $V_{N-1}$ DHF potential, with the hole
in the $(n-1)p_{3/2}$ core orbital~\cite{Vnm1_ph}.

We separate the  sum $S_I^2$ over the core excited states into
two contributions
\[
S_I^2 = \left(  S_{I}^{2}\right)_{\mathrm{core}} +
\left(  S_{I}^{2}\right)_{\mathrm{cntr}} \, .
\]
The first term is associated with the dynamic polarizability  
$\alpha_g(i\omega)$ of the closed-shell ground state $\,^1\!S_0$ and 
the second term is a corrective counter term.
\begin{equation}
\left(  S_{I}^{2}\right)_{\mathrm{core}}(i \omega)  =\left(  -1\right)
^{I+J+1}\frac{[I]}{2}\alpha_{g}\left( i \omega\right)  \, .
\end{equation}
We use the semiempirical dynamic polarizabilities for the ground
states of noble-gas atoms of Kumar and Meath~\cite{KumarMeath85}. 
The estimated uncertainty
of these core polarizabilities is less than 1\%. The primary
role of the core polarizability is to provide
the correct limit Eq.~(\ref{Eqn_TRK}) at $\omega \rightarrow \infty$.
The relative importance of the core-excitation contribution
increases for heavier systems; for example, in a similar
calculation for Fr~\cite{DereviankoC6_PRL99}, 
core excitations  contribute 23\% of 
the $C_6$ dispersion coefficient. 
The high-frequency limit, Eq.~(\ref{Eqn_TRK}),  
is accurately reproduced
by the present {\em total} reduced dynamic polarizabilities $S_I(i\omega)$.
We obtain
for Ne 9.98, for Ar 17.95, for Kr 35.95, and for Xe 53.96 compared
to the nonrelativistically exact values 10, 18, 36, and 54 respectively.

In the  $ns(3/2)_2 = [((n-1)p_{3/2})^{-1} ns_{1/2}]_2$
state core excitations to the occupied  magnetic sub-states of the
$ns_{1/2}$ particle state are not allowed by the Pauli 
exclusion principle, and neither are the core excitations
from the empty magnetic sub-state  of the hole  $(n-1)p_{3/2}$.
To remove these transitions from the core polarizability contribution
$\left(  S_{I}^{2}\right)_{\mathrm{core}}$,
we introduce a counter term $\left(  S_{I}^{2}\right)_{\mathrm{cntr}}$.  
Explicitly in the independent-electron model
\begin{eqnarray*}
\left(  S_{I}^{2}\right)_{\mathrm{cntr}} (i \omega) &=& 
[I][J]\left(  -1\right)  ^{I+J} \left(
\sum_{a}\frac{(\varepsilon_{v}-\varepsilon_{a})\langle a||d||v\rangle^2}
{(\varepsilon_{v}-\varepsilon_{a})^2 + \omega^2}  \left\{
\begin{array}
[c]{ccc}%
1 & J & I\\
j_h & j_a & j_v
\end{array}
\right\}^{2} \right. \\ 
&+& \left. \sum_{m}\frac{(\varepsilon_{m}-\varepsilon_{h})\langle h||d||m\rangle^2}
{(\varepsilon_{m}-\varepsilon_{h})^2 + \omega^2}  \left\{
\begin{array}
[c]{ccc}%
1 & J & I\\
j_v & j_m & j_h
\end{array}
\right\} ^{2} \right) \, .
\end{eqnarray*}
We estimate  the small counter-term using the Dirac-Hartree-Fock approximation.

The largest contribution to the sums $S_I$ arises from
the intermediate states in the lowest $np$ fine-structure multiplet.
The determination of electric-dipole matrix elements involved 
in the sum $\left(S_I^1\right)_{np}$ requires a knowledge
of both decay rates and branching ratios in the manifold.
The relevant lifetimes have been measured to within an error less than 1\% 
for
Ne~\cite{Kandela84,Hartmetz},  Ar~\cite{Volz98} and 
Kr~\cite{SchmoranzerVolz_Kr93}, 
and less than 3\% for Xe~\cite{Inoue_Xe_JCP84,Husson_OptCom72,Allen_JOSA69}. However, the branching 
ratios $B$ are not established  to the same precision. 
The most accurate measurements of $B$ 
in Ne~\cite{Kandela84,Hartmetz}, have an error bar of approximately 4-5\%, which 
would  introduce an uncertainty of 4-5\% in the static 
polarizabilities, and  8-10\% inaccuracy in the values of $C_6$.
To reduce the consequent errors, the experimental values of the static polarizability,
accurate to 2\%, were chosen as the reference data.

The branching ratios of transitions to
the $ns(3/2)_2$ state have been adjusted as follows.
The sum $S_3(0)$ includes only one intermediate state in the $np$ manifold,
$np(5/2)_3$,  and very small {\em ab initio} corrections. 
The $np(5/2)_3$  state has a single decay channel, so that 
the sum $S_3(0)$ is known with the experimental precision of the decay rate.
The sums $S_2(0)$ and $S_1(0)$ can be deduced from the experimental
values of the static tensor polarizability as
\begin{eqnarray*} 
S_1(0) &=& - \frac{9}{14} S_3(0) + 5\alpha_{zz}(1) - \frac{5}{4} \,\alpha_{zz}(2) \, \\
S_2(0) &=&  \frac{5}{14} S_3(0) - \frac{15}{4}\alpha_{zz}(2) \, .
\end{eqnarray*}
Removing small {\em ab initio} and semiempirical core-excitation contributions from these sums,
the sums $\left(S^1_1\right)_{np}(0)$ and $\left(S^1_2\right)_{np}(0)$ are
obtained. The branching ratios $B$  for
four states involved in the $J=1$ sum and three states in the $J=2$ sum were multiplied 
by a uniform scaling factor. 
Branching ratios for Ne~\cite{Kandela84} for the $J=1$ levels 
were multiplied by 1.0035,
and for the $J=2$ level by 0.905 in order to reproduce the experimental values 
of the static tensor polarizabilities. 
We modified the recommended values of $B$
for Ar~\cite{Wiese89} by multiplying the branching ratios of the  $J=1$
states by 0.973 and of the $J=2$ states by 0.965; the values used in the calculations
are listed in Table~\ref{Tbl_lifetimes}. 
For Kr the velocity-gauge branching ratios,
tabulated in Ref.~\cite{Doery_sp_NeKr_98} from calculations 
by Aymar and Coulombe~\cite{Aymar_78},
were multiplied by  1.127
for the $J=1$ states and by 1.0016 for the  $J=2$ states. 
For Xe, velocity-gauge values of $B$ calculated
in Ref.~\cite{Aymar_78} were multiplied by 0.927 for the $J=1$ states and 
by 0.929 for the $J=2$ states. The adjusted data for Ar and Xe 
are listed in  Table~\ref{Tbl_lifetimes}. 
Doery {\em et al.}~\cite{Doery_sp_NeKr_98} have compiled the input 
data for Ne and Kr, which have to be similarly modified.

We employ the constructed reduced polarizabilities $S_I(i\omega)$ to 
calculate the intermediate uncoupled dispersion coefficients ${C}_6^{J_a J_b}$
by quadrature using Eq.~(\ref{Eqn_C6int}). 
The coefficients are
listed in Table~\ref{Tbl_intC6}. They are to be used if the entire
molecular Hamiltonian, including quadrupole-quadrupole and higher multipoles or 
perturbation-theory orders
is to be diagonalized.
Finally, the molecular terms are obtained by the diagonalization
of $H_{\mathrm{eff}}^{\left( 2\right) }$, given by Eq.~(\ref{Heff2_Eqn}). 
Neglect of the small corrections due to the quadrupole-quadrupole interaction 
results in parameterization of term energies in the form 
\[
U(R) = 2{\mathcal E}^* -C_6/R^6 \, . 
\]
The calculated dispersion coefficients $C_6$ for various molecular symmetries 
are listed  in Table~\ref{Tbl_C6}. Since the
region close to $\omega=0$ contributes the most to the values of the
integral in Eq.~(\ref{Eqn_C6int}),
the uncertainty in the values of $C_6$ is
approximately 4\%, reflecting the 2\% experimental error in 
the static dipole tensor polarizabilities~\cite{Molof74}.
The values of the $C_6$ coefficients grow monotonically  from Ne to Xe,
due to the reduction in the energy separations between the
metastable states and the $np$-manifold.  
For heavier systems the  anisotropy in $C_6$,
arising from relativistic  effects
becomes increasingly marked, 
from 6.5\% in Ne to 16\% in Xe.

Long-range dispersion coefficients for two interacting metastable Ne atoms were 
evaluated recently by Doery {\em et al.}~\cite{Doery98_ssC6}. 
The $C_6$ coefficients were calculated 
from the diagonalization of the molecular dipole-dipole Hamiltonian
in the model space containing the lowest $np$-manifold, so limiting  the 
intermediate states to the lowest $np$-manifold states in the present formulation.
Experimental values of decay rates and branching ratios were used to 
deduce the electric-dipole matrix elements. The precision
of the calculated values of $C_6$ is about 8-10\% due to the large uncertainty 
in the branching ratios. 
The values of
$C_6$ from Ref.~\cite{Doery98_ssC6} 
for different molecular symmetries vary between 1951 and 1956 a.u.,
exhibiting much less anisotropy than  the present
results which range between  1877 and 1999 a.u.. 
The difference can be traced to the
anisotropy in the static dipole polarizabilities. Indeed, utilizing input
data from Ref.~\cite{Doery98_ssC6} we obtain $\alpha_{zz}(M=1)=192$ and
$\alpha_{zz}(M=2)=189$ a.u. if we include only 
the $np$ manifold as in Ref.~\cite{Doery98_ssC6}. 
While  $\alpha_{zz}(M=1)$ agrees with the experimental value~\cite{Molof74} 192(4),
the $\alpha_{zz}(M=2)$ is overestimated  by three standard deviations compared to the
experimental value 180(3) a.u..

The accuracy  of the dispersion coefficients could be improved by 
applying relativistic all-order many-body 
methods~\cite{Vnm1_ph,Avgoustoglou_PRA98} to calculate transition amplitudes 
between $ns-np$ manifolds. Such {\em ab initio} calculations
are intrinsically more challenging than for alkali-metal atoms;
the accurate experimental lifetimes would provide an excellent gauge of accuracy.  

Our values of $C_6$ coefficients will be useful in studies of cold collisions
of metastable rare-gas 
atoms~\cite{Orzel_Xe_PRA99,Doery_sp_NeKr_98,Doery98_ssC6,Optical_Control_Xe_PRL95}.
For example, we can estimate the rate coefficient for Penning ionization
by ignoring spin-polarization and assuming that every trajectory that surmounts the angular momentum
barrier leads to ionization~\cite{Orzel_Xe_PRA99,Bell_JPB68}. 
The corresponding rate coefficient is given by~\cite{Bell_JPB68}
\[
 k= 6.35 \times 10^{-9} \frac{C_6^{1/3} T^{1/6}}{\mu^{1/2}} \, {\mathrm cm^3 \; s^{-1}}
\, ,
\]
where $\mu$ is the reduced mass measured in units of the electron mass and 
$T$ is the temperature.
Combined with short-range potentials~\cite{Doery98_ssC6,Kotochigova_PRA2000}
a number of other properties could be determined. 
For example, scattering lengths of elastic collisions could be found,
providing input for mean-field equations describing dilute quantum gases. 

This work was supported  by the U.S. Department of Energy,
Division of Chemical Sciences, Office of Energy Research. Thanks are
due to M. R. Doery, S. Kotochigova, and J. F. Babb for useful discussions. 
The authors are grateful
to W. R. Johnson for providing the B-spline routine for the $V_{N-1}$ 
Dirac-Hartree-Fock potential.

\begin{table}[h]
\begin{center}
\caption{ Input data for Ar and Xe calculations. 
Lifetimes $\tau$ for Ar are from Volz and Schmoranzer, Ref.~\protect\cite{Volz98},
and for Xe from Inoue {\em et al.}, Ref.~\protect\cite{Inoue_Xe_JCP84},
except where noted. Branching ratios $B$ to the $ns(3/2)_2$ level are adjusted to reproduce
experimental tensor dipole polarizabilities, as discussed in the text.
\label{Tbl_lifetimes}}
\begin{tabular}{llrlr}
&\multicolumn{2}{c}{Ar, $n=4$} & \multicolumn{2}{c}{Xe, $n=6$}  \\
state  &  $\tau$, ns & $B(\%)$  & $\tau$, ns & $B(\%)$ \\
\hline 
$np'(1/2)_1$ & 27.85(7)& 17.66   & 43.5(1.5)\tablenotemark[2] 
		    		     & 5.15    \\
$np'(3/2)_2$  & 29.01(7)& 10.59  & 38.1(1.3)\tablenotemark[3] 
             			     & 2.30    \\
$np'(3/2)_1$  & 29.83(8)&  1.84   & 49(2)     & 1.16	\\
$np(3/2)_2$   & 28.52(7)&  68.55  & 31(1)     & 65.42\\
$np(3/2)_1$   & 29.62(7)&  14.97  & 37(1)     & 7.45	 \\
$np(5/2)_2$   & 31.17(7)&  27.76  & 39(1)     & 33.29  \\
$np(5/2)_3$   & 29.00(7)&  100.00 & 31(1)     & 100.00  \\
$np(1/2)_1$   & 39.2(2.2)\tablenotemark[1]
                                    & 72.11  &  38(1)     & 87.60	   
\end{tabular}         
\end{center}
\tablenotetext[1]{Wiese {\em et al.}, Ref.~\cite{Wiese89}.}
\tablenotetext[2]{Allen {\em et al.}, Ref.~\cite{Allen_JOSA69}. }
\tablenotetext[3]{Husson and Margerie, Ref.~\cite{Husson_OptCom72}. }
\end{table}

\begin{table}[h]
\begin{center}
\caption{ Intermediate dispersion coefficients ${C}_6^{J_a J_b}$, a.u., 
multiplied by a factor $10^{-3}$. \label{Tbl_intC6} }
\begin{tabular}{ccccccc}
   & $C_6^{11}$ & $C_6^{21}$ & $C_6^{22}$ & $C_6^{31}$ &  $C_6^{32}$ & $C_6^{33}$\\\hline
Ne & 4.945     &  -7.333    &	10.89	&   10.92   &  -16.21	  &  24.14\\
Ar & 12.48     &  -17.42    &	24.39	&   27.12   &  -37.88	  &  59.02\\
Kr & 15.12     &  -20.18    &	27.01	&   31.80   &  -42.44	  &  66.96\\
Xe & 23.24     &  -29.22    &	36.91	&   47.74   &  -60.06	  &  98.31\\
\end{tabular}
\end{center}
\end{table}

\begin{table}[h]
\begin{center}
\caption{ Dispersion coefficients $C_6$ in a.u.
for the interaction of two metastable $ns(3/2)_2$ noble-gas atoms. 
\label{Tbl_C6}}
\begin{tabular}{crrrr}
Term      &   Ne &  Ar  &  Kr & Xe    \\
\hline\\[-10pt]
 $4_g$    & 1877&4417 & 4994& 7138 \\[2pt]
 $3_g$    & 1919&4565 & 5195& 7490\\
 $3_u$    & 1922&4583 & 5224& 7557\\[2pt]
 $2_g$    & 1967&4751 & 5459& 7991\\
 $2_g$    & 1935&4629 & 5286& 7664\\
 $2_u$    & 1934&4623 & 5276& 7641\\[2pt]
 $1_u$    & 1983&4811 & 5543& 8148\\
 $1_g$    & 1982&4810 & 5541& 8145\\
 $1_g$    & 1920&4574 & 5210& 7524\\
 $1_u$    & 1920&4574 & 5210& 7526\\[2pt]
 $0_g^{+}$& 1999&4872 & 5629& 8311 \\
 $0_g^{+}$& 1968&4756 & 5467& 8010 \\
 $0_u^{-}$& 1966&4747 & 5452& 7975 \\
 $0_u^{-}$& 1877&4418 & 4996& 7140 \\
 $0_g^{+}$& 1877&4418 & 4996& 7140  \\
\end{tabular}         
\end{center}
\end{table}
\newpage


\end{document}